# Usability Evaluation of Learning Management Systems in Sri Lankan Universities

Thuseethan, S.[1], Achchuthan, S.[2], Kuhanesan, S.[3]

[1,2]Sabaragamuwa University of Sri Lanka, Sri Lanka
[3]Vavuniya Campus of the University of Jaffna, Sri Lanka
[1]thuseethan@gmail.com
[2]achchu2009@gmail.com
[3]kuhan9@yahoo.com

**Abstract**
As far as Learning Management System is concerned, it offers an integrated platform for educational materials, distribution and management of learning as well as accessibility by a range of users including teachers, learners and content makers especially for distance learning. Usability evaluation is considered as one approach to assess the efficiency of e-Learning systems. It is used to evaluate how well technology and tools are working for users. There are some factors contributing as major reasons why the LMS is not used effectively and regularly. Learning Management Systems, as major part of e-Learning systems, can benefit from usability research to evaluate the LMS ease of use and satisfaction among its handlers. Many academic institutions worldwide prefer using their own customized Learning Management Systems; this is the case with Moodle, an open source LMS platform designed and operated by most of the universities in Sri Lanka. This paper gives an overview of Learning Management Systems used in Sri Lankan universities, and evaluates its usability using some pre-defined usability standards. In addition it measures the effectiveness of LMS by testing the Learning Management Systems. The findings and result of this study as well as the testing are discussed and presented.

## 1. INTRODUCTION

E-learning has a well-developed approach to the creation and sequencing of content-based, single learner, self-paced learning objects (Dalziel, 2003). Open learning is defined as a student-centered approach for education that eliminates all barriers to access while providing a high degree of learner autonomy (Maxwell, 1995). Nowadays the way of delivering a course of study through some electronic media is dramatically increased. Here in this way of delivery the majority of communication between teachers and students occurs in non-continuous fashion. Computer based systems increase the efficiency and reduces the risks involved in any mode of activity (Thuseethan, 2014). Further in technologically mediated educational process, an efficient two-way communication between teachers and students is extremely important. During the last ten years, many universities and higher educational institutions have started offering distance education courses for their on-campus students because of the following reasons (Aybay et al., 2002).

- Online course development: The University gains more experience on it
- Establishment: Gains more experience on the management of online programs and this perhaps leads to the establishment of an institute
- Quick response from the students involving online courses respond quickly
- Staff development: Train sufficient number of teaching staff who are qualified in evolving distance educations

Most of the modern institution providing higher education desires a Learning Management System (LMS) to handle teaching and learning activities. Somehow it is essential to offer electronic lecture materials for students to access via the internet anywhere at any time. Bearing in mind the significance of all these needs, and believing that distance education will become more important in the education system, all universities in Sri Lanka initiate the practice of learning management systems. Learning management systems are essential for content development and management of online programs (Aybay et al., 2002). One of the most important features of LMS is to provide an environment for learning and teaching without the restrictions of time or distance (Epping, 2010). Most of the researches concentrate on performing comparative or evaluation studies on learning management system technical or pedagogical issues. Even though a few number of researches have been done by concern these systems accessibility or usability. In this sense usability is one of the major term in Human-Computer Interaction, defined as the ease with which a user can learn to operate, prepare inputs for, and interpret outputs of a system or component [IEEE Std.610.12-1990]. In the context of Learning Management System usability testing concentrate on learning about the understanding of the user engaged in it.

Due to the complexity of human nature and individual differences, objective and systematic assessment of human behavior and performance remains highly difficult (Bellotti et al., 2013). But conducting usability evaluations have been taken as a crucial quality assessment technique in evaluating learning management systems. Numerous usability evaluation methods have been developed and materialized in research and practice in the field of usability engineering. Presently, usability is becoming a significant concern for e-learning and for learning management systems development and most practitioners perceive usability as a crucial factor in e-learning applications planning and usage (Inversini et al., 2006). Evaluating the usefulness and effectiveness of learning management system can benefit both academic institution and students as well.

In this paper we discuss on the findings of usability evaluation in Sri Lankan Universities and deliberate their implications.

## 2. LITERATURE REVIEW

### 2.1. Learning Management Systems in Sri Lankan Universities

The rapid development of ICT infrastructures in Sri Lanka motivates every educational institution to make use of the internet as a medium of communication among the students. The effective and efficient access to learning materials achieved by the concepts and methodologies of technology-based learning. Increasing use of e-learning materials becomes a crucial resource for institutions. LMS has been widely used in higher education due to various advantages including flexible learning times and boundless distance education (Hamuy et al., 2009).

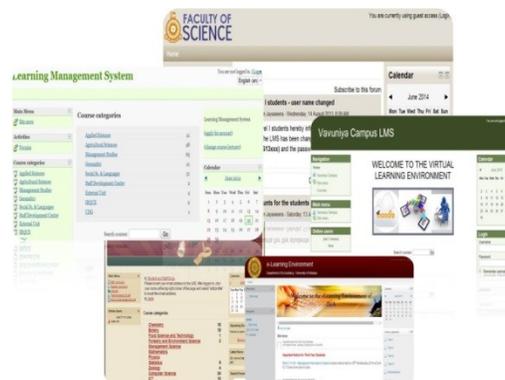

*Figure 1: Sample LMS home pages in Sri Lankan Universities (Courtesy: All universities in Sri Lanka)*

In most Sri Lankan state universities, Moodle open source platform is used as LMS. Figure 1 shows some of the learning management system interfaces of Sri Lankan state universities which are currently in use. The universities in Sri Lanka are expecting at-least the following functionalities from well-organized Moodle learning management systems:
- The registration of teachers and students in the educational portal
- Planning and scheduling the course and the way of structuring it
- Provide the way of delivery or making the course accessible for registered users
- Track the students' progress as well as producing automatic reports of students' performance
- Communicating students with each other through forums, mails, file sharing and chatting applications
- Teacher and student evaluation

## 2.2. Why Moodle in Sri Lanka

Modular Object Oriented term Developmental Learning Environment (Moodle) is a course management system through the Internet, also recognized as a Learning Management System (LMS) or a Virtual Learning Environment (VLE). It is a free web learning environment that educators can use to model effective online learning platforms. In this sense, it can be used to model effective online learning programs. One of the major advantage is it is an open source, which can be used by any users, modifying with programming knowledge and adapt the environment according to their own desires. It can be installed at any number of servers without any cost and there is no maintenance costs required to pay for upgradings. This learning platform has worldwide users such as universities, societies, schools, teachers, courses, instructors and even in businessmen. Likewise Sri Lankan universities also adapt to this. The design of Moodle is totally based on socio-constructivist pedagogy (Brandl, 2005). This means its goal is to give a set of tools that backing an inquiry- and discovery-based approach to online learning process.

The great success of Moodle is due to the fact that it satisfies the guidelines for best LMS. The best LMS solution is defined in this study as one in which all LMS components are considered within the total learning infrastructure of universities such that maximum student success is ensured from both an institutional and System perspective (Randal, 2010). Aspects of these components in terms of students' perspective success were assessed by the following attributes:
- Interoperability and Flexibility
- Cost effectiveness
- Support and Training
- Ease of Use
- Scalability
- Sustainability

In reality, for instance Moodle gives a more sophisticated and structured environment. It looks more like a set of tools that share an efficient learning environment. These are some strong reasons behind the wide range of usage in Sri Lankan context.

## 2.3. Usability and Learning Management Systems

Web usability arose as research field at the very beginning of the Internet era (Rukshan et al., 2011). To enhance the usability of learning management systems, human computer interaction holds a major role in attaining the goal of improving user performance (Sung et al., 2012). Many past researches in human computer interaction have offered beneficial information on how users fit to perform and think about the system to use it easily. Research in this area offers significant insight for technology usability and consideration of the user for the design element of human computer interaction (De Lera et al., 2010). Based on the International Organization for Standardization, the term usability refers that users can effectively use a tool or system to accomplish a task with satisfaction and ease (ISO 9241-11, 1998). In user's perspective, the use of Learning Management System is constrained by the human's perceptual and cognitive abilities (Thuseethan et al., 2014).The better human computer interaction that offers the learning management systems users, the easier of use and greater satisfaction users will have within systems or tools they involved. Usability can improve the learning experience for students (Tselios et al., 2008) as well as academic performance. Therefore, a sensible design of human computer interaction with usability study is one of the crucial components in the design and development of learning management systems.

Based on the Shackel's proposal usability can be viewed in terms of four major operational criteria, those are effectiveness; learn ability, flexibility and attitude. This study involves the testing of all four operational criteria on learning management system.
- Effectiveness – The performance in accomplishment of tasks by some percentage of the users within the system
- Learn ability – The degree of learning to accomplish tasks. Learning also includes the time taken to learn and relearn the system.
- Flexibility – The adaptation to variation in tasks and environments which can be accommodated by the design.
- Attitude – The user satisfaction with the system whether to continue use the system or enhance their use of the system

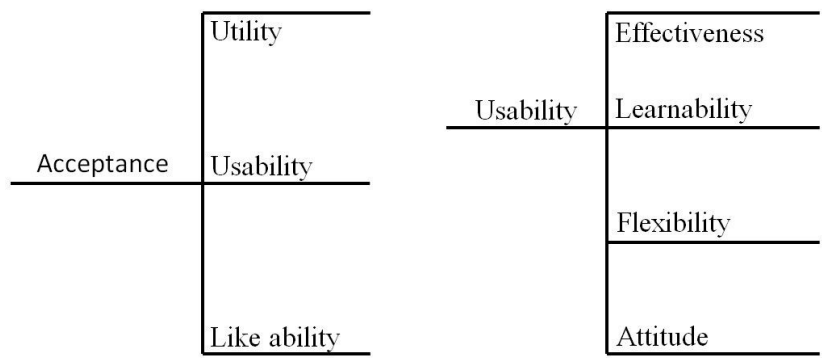

*Figure 2: Shackel's definition of Usability (1991)*

## 3. EVALUATION METHODOLOGIES

We used different approaches to do the usability evaluation. Most studies on learning management system focus on the technical parts of the systems. These kinds of studies are rarely assessing the effectiveness and user satisfaction in learning management systems in terms of users. The techniques used for evaluating the usability of learning management systems have varied from simple checklists to more complicated standardized questionnaires. Many research studies have been conducted to evaluate the usability of existing learning management systems. Selection of right technique for evaluation depends on the complexity and functionality of the learning management systems and sometimes on goal that system has.

### 3.1. Approach 1
The purpose of this approach is to present some first findings of the usability of learning management systems among a selected group of students with advanced computer proficiency. This study took place in seven different universities in Sri Lanka and more than two hundred students taken from computer science based departments to answer the evaluation questionnaires. We did this survey during the last few day of semester. Because in last few days of the semester the usage of learning management is high comparing with normal days, during that time students used to submit the assignments, ask questions and clearing their doubts in the discussion boards, download course materials and handouts, check notices and complete online quizzes.

In this approach we used two standard questionnaires for the evaluation. In both questionnaires five-scale Likert scale (Strongly Agree {4}, Strongly Disagree {0}) were used to mark the students' response:
1) The SUS (System Usability Scale) (see Appendix A), a mature questionnaire constructed by John Brooke in 1986. This questionnaire comprises 10 statements and it is very robust and has been widely used and adapted to evaluate usability.
2) Other new questionnaire was constructed based on two standard questionnaires which are: Usability and User Satisfaction Questionnaire (Zins et al., 2004) and the Web-based Learning Environment Instrument (Chang, 1999). The newly generated questionnaire LMS Usability Questionnaire consisted of 10 questions picked from both questionnaires.

Questions picked from these two standard questionnaires asses the following areas of usability in learning management systems (see Appendix B):
- System layout design
- System functionality
- Ease of use
- Learnability
- Satisfaction
- Outcome/future use
- System usefulness.

### 3.2. Approach 2

This approach involves the testing on the effectiveness of the learning management system as the major study. During this phase, the candidates are given with tasks list and questionnaire to observe the responses. Defined task list is translated into scenarios based activities with some specific goals. Based on the Shackel's (1991) four factors on usability four questions were used to evaluate usability. Figure 3 shows the research methodology framework used to evaluate the usability of learning management system which is classified into four factors in the areas of study.

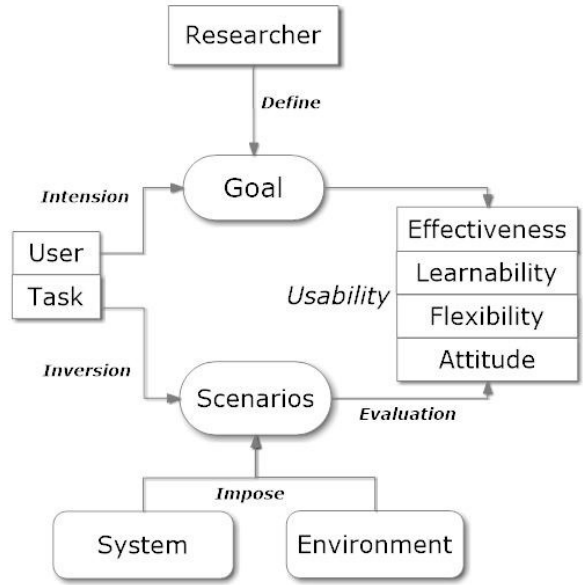

*Figure 3: The Research Theoretical Framework*

Definition of the goal is done by the researcher which has been intended by users. The user accomplishes the task by doing the inverted scenarios; one single task structured into one or many scenarios. Scenarios depend on the system and environmental state, where the system is the computer system and environment comprises physical aspects such as proper heating, lighting, layout, operating conditions as well as psychological facets such as the provision of help, training, customer care and socio-political features such as the organizational environment in which the interaction happens. Finally the relationship between usability of desired goal and achieved goal get compared and analyzed. The acceptance of learning management systems is measured by the usability factors such as effectiveness, flexibility, learnability and attitude in particular environment and system. Real-time evaluation is probably one of the most demanding types of evaluation practice, requiring not only a wide range of skills from evaluators but also a tightly focused professional approach in order to meet the time demands (Clarke et al., 1997). This testing approach involves students or users of the learning management system to work on typical tasks using the real system and in real time. In real-time evaluation of learning management systems, four major tasks were formulated by dividing those into sub tasks based on the three main features or functions of the system (see Appendix C). All the tasks should be completed within fifteen minutes. After completing every section of the task the subject has to give comments. Real time results can be used by designers to make changes on the system design (Genise, 2002; Sriharan, 2014). The final result from this real-time test can be used to illustrate how the user interface, speed, quality and the overall of the learning management system can supports the users in their learning process.

## 4. RESULTS AND DISCUSSIONS

Figure 4 displays the overall response of 201 students to each question in SUS questionnaire as average response which varies from 0 to 4. The average students' response to the positive statements 1,3,5,7 and 9 were above midpoint which means that the students found the current leaning management systems easy to use and its functionalities were designed properly and well integrated. In the meantime, the responses to the negative statements 2, 4, 6 and 8 discovered that even though the current learning management was user friendly and easy to use, it still has some inconsistency, complexity and irregular actions in its functionalities.

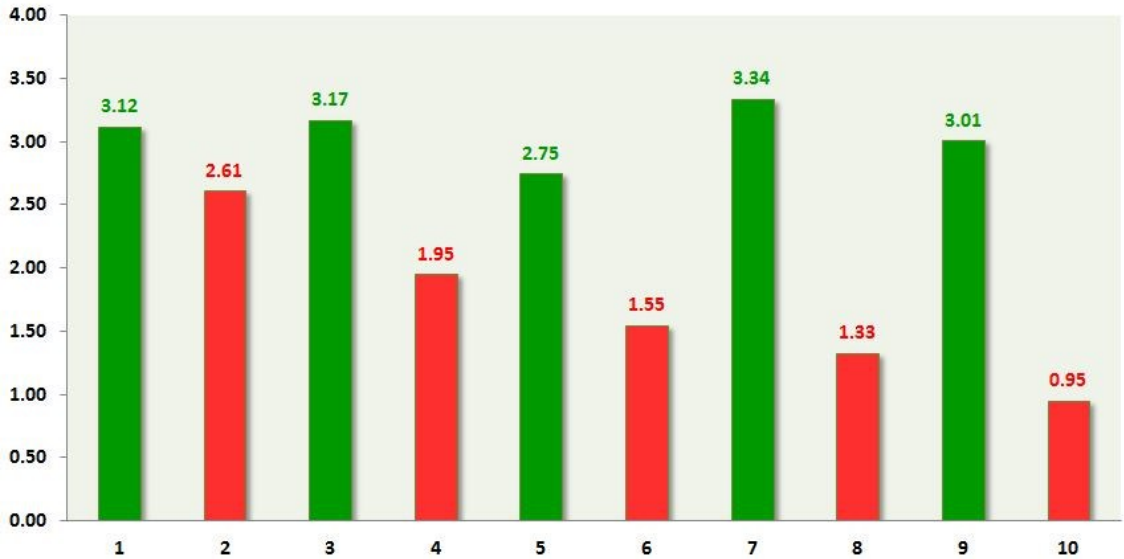
*Figure 4: Average score of students' response for each question in SUS*

Further we analyzed the users respond to negative questions positively to find the reason behind their response. By considering the response to statement number 2 shows that even though they like to use present learning management systems they found some kind of complexity while using it. Another important negative statement is number 4, even though most of the students were familiar with using computers, web and other information systems, yet some of them found the learnability of the system was in the border line and require help from specialized person.

On the other hand next questionnaire consisted of 10 questions picked from both the Usability and User Satisfaction Questionnaire and the Web-based Learning Environment Instrument. The result is somehow related to the findings of SUS questionnaire, however with more insights about the level of satisfaction practiced in learning management systems. Figure 5 shows the overall response of the students to the selected statements from the Usability and User Satisfaction Questionnaire. Responses to statements other than 1, 5, 6 and 8 were above midpoint. According to those four low response statements most of the users found problems in interfaces, appropriate error messages, recovery mechanisms and location of online materials. Apart from this we must comment that most students indicate some significant functional and technical issues in it.

- The malfunction of the search feature
- The post organization in the forum and discussion board
- The inconsistency in downloading course materials

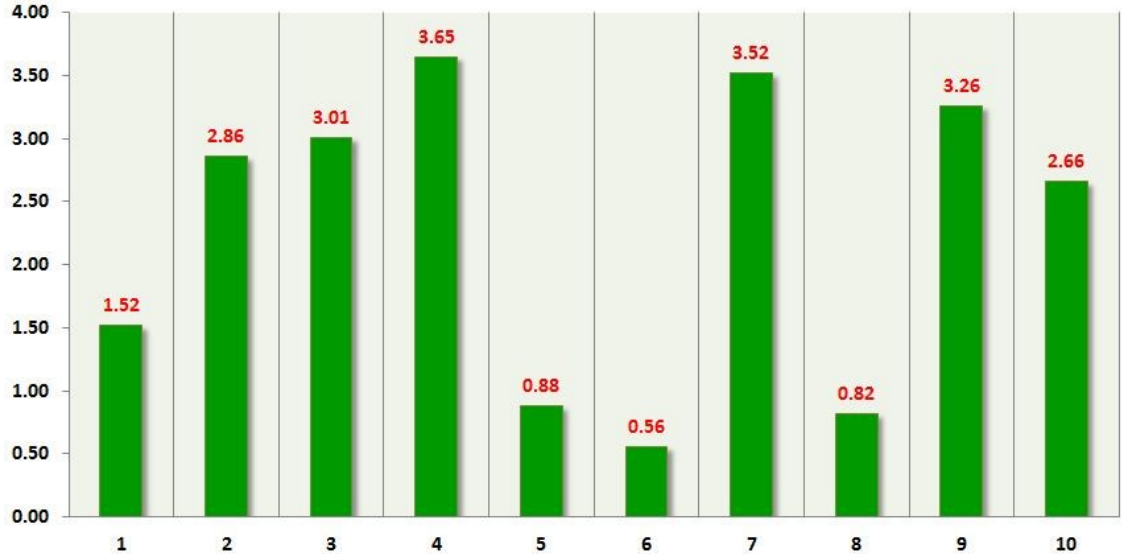

*Figure 5: Average score of students' response for each question in combined questionnaire*

According to the user feedback we found some major issues in present learning management systems with evidence.

### 4.1. Lack in First Impression

Most of the users have reported about the bad login user interface. For the very first time users are overloaded with information when logging into learning management systems. At that time they lose their focus on goal. Some information could be omitted in the first time use such as old and read news, course details and e-mail messages. Users also demanded the ability to maximize each sub window on the welcome page, in addition to a search function. Some users prefer search function as a crucial means of navigation. Figure 6 shows one evident for bad login design.

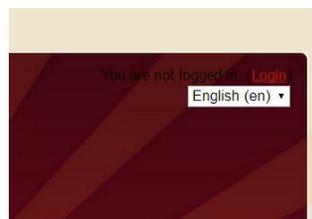

*Figure 6: Evidence for bad login design reported by users*

### 4.2. Visual inconsistencies

Moodle allows the teachers' or source creators to integrate own material which is likely to result in visual inconsistencies because of different styles they used. One teacher appends with one visual style and the other teacher is likely to have deployed a totally different style. Therefore, in this case two or more different styles lead to inconsistencies in learning management systems. This inconsistency is not a major issue but it does, however, give the user the impression of chaos and lack of professionalism in design. Figure 7 shows the usage of different size of fonts in present learning management systems.

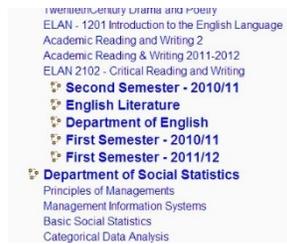
*Figure 7: Evidence for usage of fonts in different sizes for same category*

Figure 8 indicates that the maintainers of present learning management systems have been sloppy in designing hyperlinks. A well-structured hyperlink is characterized by the fact that the system user immediately knows to click on the link without time delay. It is a convention that hyperlinks should be underlined and preferably be in blue colour. But in present learning management systems one is blue coloured, one is black coloured and one is in grey. Some are hyperlinked and remaining is not underlined. The maintainers should have reduced the number of hyperlink styles to one unique style to maintain consistency.

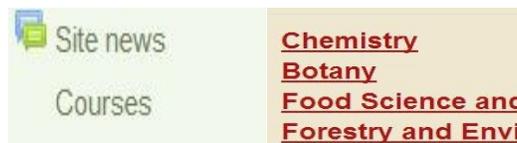
*Figure 8: Evidence for different kinds of hyperlinks*

Another main problem is colour usage in interfaces. Use few colours would provide the user interface a solider sense of consistency and uniformity in look and style. The combination of green and blue (analogous colors) were used in some present learning management systems. Red and orange is also used (complementary colors). In some parts yellow is also used. Red, yellow and blue encompass a complex triadic color scheme. The students found it tough to group these overlapping color schemes. The users would probably perceive the system as more consistent if fewer colors were used. The lack of consistency not only creates a problem in accessing information in these very complex systems but also increases operational and training costs to the users.

Another major issue is each course instructor is responsible for configuring the menu, title, backgrounds, fonts and the folder structure for each course. A better solution would probably be a standardized menu, title, backgrounds and folder structures. A standardized structure would allow students to more easily orient themselves and reuse their knowledge from one course to another without retraining. Most of the learning management systems support one locale which is English. The internationalization capability in presenting one does not match the rapid increase in internationalization at universities.

### 4.3. Lack of error prevention and recovery

Reliable operation of a computing system depends on both error detection and error recovery (Horning et al., 1974). Some users found some appropriate presentation of error messages. Figure 9 shows one evident for bad error presentation without colour or warning sound. Some users reported that there are insufficient back buttons in interfaces. Sometimes in some interfaces there is unnecessary placement of back buttons.

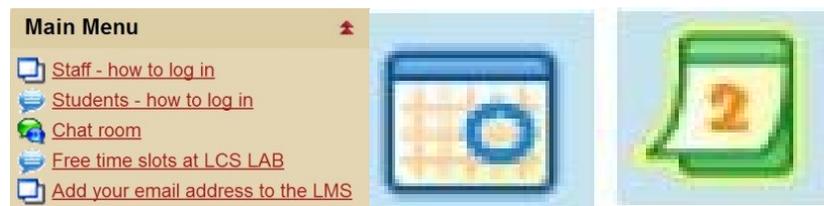

*Figure 9: Evidence for bad error presentation*

### 4.4. Icon recognition or eye candy

A pictogram is a stylized figurative drawing that is used to convey information of an analogical or figurative nature directly to indicate an object or to express an idea (Charles et al., 2007). Icons are widely used in present learning management systems; it's an enhanced way of presentation than text. Icons can be used together with text as a visual substitute to do tasks. Because the reading is cognitively more challenge than well-integrated icons. Users found two major mistakes as use of same icons for different operations and use of different icons for same operation in different interfaces.

*Figure 10: Evidence for inconsistency in using icons*

Figure 10 shows the inconsistency in using icons. Same icon used for students' login help and free time slot option. On the other hand two different icons used for calendar option in different interfaces.

## 5. CONCLUSION AND FUTURE WORK

The outcome of this work indicates the overall level of the effectiveness of learning management system constructed in students' perspective. The results found that most of the students liked present system and find it very easy to access. However, it suffers from some functional, design and technical problems in its usability. Further some of the major findings through this study are 1) It is useful that the system is trying to do much more than is required by user 2) Currently it is hard to use some important functions like login and assignment submission 3) Teachers should be given with proper guidelines or less freedom while uploading or organizing the system 4) Maintainers are not efficient and not maintaining the components according to HCI standards. We conclude that each and every revision of present systems should be undergone or proofread by an expert or central authority to maintain the consistency.

Since this research is a preliminary stage study on learning management system, it is supposed that it provides some awareness into the usability of current system. Furthermore usability studies can be lead to evaluate adapting other existing usability evaluation techniques. In future usability studies can be conducted to refine the existing HCI standards through users' feedback and further virtual reality can be included inside the current system.

**Appendix A**

List of System Usability Scale (SUS) questions (John Brooke, 1986)

| No. | Question |
|---|---|
| 1 | I think that I would like to use LMS frequently |
| 2 | I found LMS unnecessarily complex |
| 3 | I thought LMS was easy to use |
| 4 | I think that I would need the support of a person with technical knowledge to be able to use LMS |
| 5 | I found the various functions in LMS were well integrated |
| 6 | I thought there was too much inconsistency in this system |
| 7 | I would imagine that most people would learn to use LMS very quickly |
| 8 | I found LMS very cumbersome to use |
| 9 | I felt very confident using LMS |
| 10 | I needed to learn a lot of things before I could get going in browsing LMS |

**Appendix B**
List of question items picked from Usability and User Satisfaction Questionnaire (Zins et al., 2004) and the Web-based Learning Environment Instrument (Chang, 1999).

| No. | Question |
|---|---|
| 1 | I liked using the interface of LMS system |
| 2 | Overall, this system was easy to use |
| 3 | It was easy to learn to use the system |
| 4 | I believe I could become productive using this system |
| 5 | The system gave error messages |
| 6 | Whenever I made a mistake using the system, I could recover easily and quickly |
| 7 | I can access the learning activities at times convenient to me |
| 8 | The online material is available at locations suitable for me |
| 9 | LMS enables me to interact with other students and the tutor asynchronously |
| 10 | I am confident in using this technology |

**Appendix C**
List of tasks inverted into scenarios to scale usability in real-time.

| Task | Scenario |
|---|---|
| 1 | Step 1. Select one of the course that you have selected<br>Step 2. Click on the quizzes button LMS<br>Step 3. Answer the test within five minutes<br>Step 4. Submit the result<br>Step 5. Comment on task 1 |
| 2 | Step 1. Click 'MyCourses' button<br>Step 2. Find a course material<br>Step 3. Display your result<br>Step 4. Search the course material within 3 minute<br>Step 5. Comment on task 2 |
| 3 | Step 1. Use the login system<br>Step 2. Change the password within 2 minute<br>Step 3. View the activity log<br>Step 4. Comment on task 3 |
| 4 | Step 1. Choose any subject<br>Step 2. Post one message asking the doubt from your tutor<br>Step 3. Send the response to one of your friend's<br>Step 4. Do both subtasks in 5 minutes<br>Step 5. Comment on task 4 |